# Delay Monitor Circuit for Sensitive Nodes in SRAM-Based FPGA

Mostafa Darvishi[1], *Student Member, IEEE*, Yves Audet[1], *Member, IEEE*, and Yves Blaquière[2], *Member, IEEE*

[1]Electrical Engineering Department, Polytechnique Montréal, QC, Canada

[2]Electrical Engineering Department, École de Technologie Supérieure (ÉTS), Montréal, QC, Canada

*Abstract*

This paper presents a novel monitor circuit architecture and experiments performed for detection of extra combinational delays in a high frequency SRAM-Based FPGA on delay sensitive nodes due to transient ionizing radiation.

**Presenter:**
Mostafa Darvishi, Ph.D. Student
Electrical Engineering Department
Polytechnique de Montreal
2900 Boulevard Edouard Montpetit,
Montréal, QC, Canada, H3T 1J4
Tel: 514 340-4711
Fax: 514 340-4147
Email: mostafa.darvishi@polymtl.ca

## I. INTRODUCTION

FIELD Programmable Gate Arrays (FPGAs) are an attractive solution to implement systems for space and aeronautic applications. Configuration bits in SRAM-based FPGA are sensitive to radiation and have been studied over years [1, 2]. Almost 98% of all memory elements in SRAM-based FPGAs are configuration bits, of which 90% and more control the routing resources [3]. Recent studies reported observations of permanent extra combinational delays due to SEU induced by proton radiation [4, 5]. Circuit-level models and simulations determined the most probable root cause of delay changes being bit flips of SRAM-cells configuring interconnections and switch boxes, which create shorts between routed and unused wires [6]. Experiments in [4] showed that Single Delay Change (SDC) due to the bombardment of the top side of the Virtex-5 FPGA with proton irradiation could be as large as 128 ps for the core circuit. Cumulative Delay Change (CDC) up to 422 ps was also observed.

These permanent delay changes may lead to malfunction of circuits by generating delay faults. Increasing the timing margin or reducing the operating frequency is a trivial mitigation technique to reduce the likelihood of these delay faults and tolerating delay changes on combinational paths. For high frequency FPGA designs where timing closure is an issue, it might not be possible to use these additional timing margins. To reduce the timing cost, some mitigation techniques employ in-situ monitors on data path to sample signals before or after the clock edge in order to predict or to detect timing errors respectively [7-10]. The TIMBER technique [7] detects a late arriving signal and correct the delay fault by borrowing time from successive pipeline stages. This technique is not timing efficient for most SRAM-based FPGA applications due to the overhead imposed by their flip-flops' topology. Other timing error detection techniques allow high frequency operation at the expense of sporadically experiencing run-time timing errors on the affected path, such as proposed in [10] tailored to network-on-chip applications. The detection and recovery mechanisms of these technique are however application dependent and not amenable to any general applications. The scan flip-flop [9] allows concurrent error detection and correction by a extra single clock cycle for error correction while stopping the whole design on error detection and re-executing erroneous operations. The large extra logic gates occupy high silicon area and create performance degradation.

The main contribution of this paper is a circuit that performs real time monitoring of single delay change (SDC) on sensitive nodes. The monitor can be configured according to the timing budget (or available timing slack) of each selected sensitive node in the FPGA. An experimental setup confirmed the functionality of the monitor circuit in real implementation. This setup emulates SEU injection into the FPGA and allows locating the exact configuration bit address and position of the fault within the configuration memory frames.

This paper is structured as follows. The motivation and methodology of SDC detection followed by a description of the proposed monitor architecture are drawn in Section II. Validation of the proposed technique with an experimental setup on a Virtex-5 FPGA is presented in Section III along with further experiments to detect cumulative delay changes (CDCs) and to locate upset configuration bit address and finally, we conclude in Section IV.

## II. MOTIVATION AND DELAY CHANGE MONITOR

An SEU affecting a configuration bit of interconnections or switch boxes in SRAM-based FPGA can permanently delay the arrival of data signal, labeled Signal Under Test (SUT), from a StartPoint flip-flop to any EndPoint flip-flop in a synchronous circuit as shown in Fig. 1. A Delay Change (DC) induced by an SEU on a SUT larger than its timing slack ($\tau_{slack}$) may lead to circuit malfunction. $\tau_{slack}$ is defined as the delay between the combinational logic data arrival time ($\tau_{CQ}+\tau_{CL}$) and the active edge of the clock ($T_{clk}$), minus the flip-flop setup time $\tau_{su}$ (Fig. 2 (b)). A data signal path with slack smaller than a DC threshold ($DC_{Th}$) can potentially generate a delay fault, and is therefore named *Sensitive node* in this paper. The likelihood of delay faults is reduced if the clock frequency is such that $\tau_{slack} > DC_{Th}$ for all the circuit nodes. However, this timing overhead could be prohibitive for high speed circuits. Proton irradiation experiments conducted on Virtex-5 FPGA showed that cumulative DC (CDC) on a signal path can be 3.3 times larger than SDC [4, 5].

In order to reduce the timing overhead of making a circuit tolerant to $DC_{Th}$, we propose to make the circuit tolerant to SDCs with a clock frequency such that $\tau_{slack} > DC_{Th}$ for all the *sensitive nodes*.

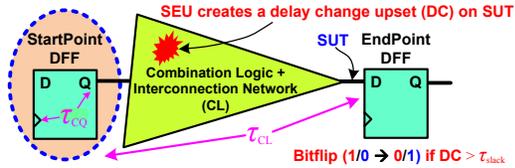

Fig. 1. Timing issues in synchronous circuits: SEU implication on a configuration bit of interconnections or switch boxes. CL stands for combinational logic.

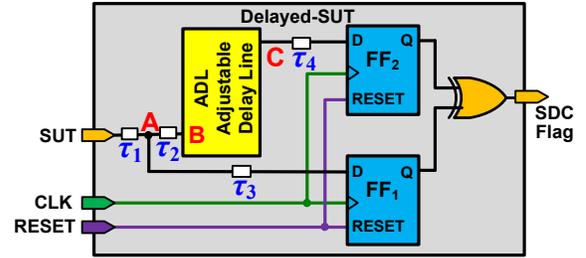

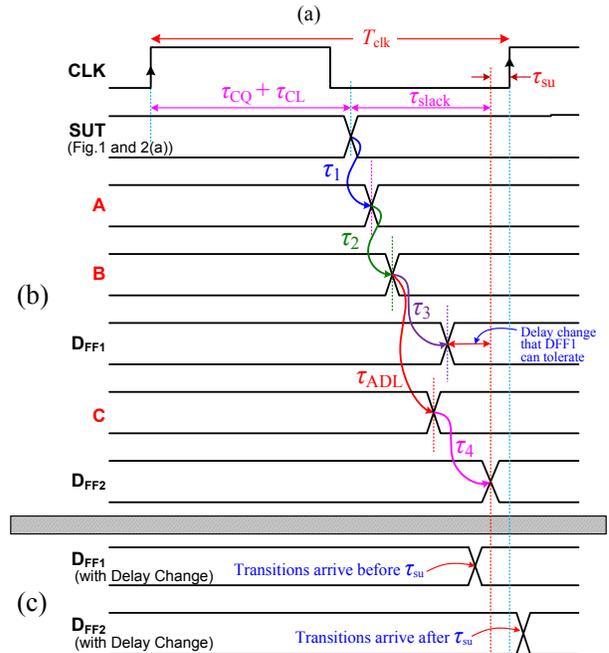

Fig. 2. (a) SUT-delayed Monitor architecture for single delay change detection. Its timing diagram is shown (b) without delay change and (c) with delay change.

SDC monitors are also added on *sensitive nodes* to detect any single delay change. Actions can be taken as soon as a SDC is detected, such as scrubbing (localized bitwise partial reconfiguration), raise a system flag, reduce the clock frequency [11], or apply any other mitigation actions. Stopping the clock signal is not the preferred solution for designs implemented into FPGAs, especially for real-time control applications. Therefore, on delay change detection, the system clock frequency could be temporarily reduced such that $\tau_{slack} > 2\,DC_{Th}$ for all sensitive nodes, in order to prevent any delay fault due to other delay changes and to avoid any system down time. This reduced frequency mode is applied until the circuit affected by an SEU is partially or entirely reconfigured and the system state is restored.

### A. Real-Time Monitor for Single Delay Change Detection

The proposed monitor for SDC detection raises a flag on the occurrence of a delay change induced by SEU on SUT in Fig. 2. Its timing diagram emphasizes routing delays $\tau_1$ to $\tau_4$ that can be significant in the monitor. An adjustable delay line (ADL) is configured with a threshold delay that satisfies the following equation:

$$\tau_{slack} \geq \tau_1 + \tau_2 + \tau_4 + \tau_{ADL} + DC_{Th} \quad (1)$$

where $DC_{Th}$ is the amount of single delay change that can be tolerated.

In normal operation without SDC (Fig. 2 (b)), the SUT transition propagates up to $D_{FF1}$ and arrives before $T_{CLK} - \tau_{su}$ and both flip-flops sample the same data value and the circuit runs without any error. When a delay change is upset, the transitions at $D_{FF1}$ and $D_{FF2}$ arrive before and after the clock transition respectively (Fig. 2 (c)), the SDC flag is asserted. Many delay rules to make the SDC monitor functional in FPGA must be respected, as presented in the following section.

Fine Adjustable Delay Lines (ADL) can be made with carry logic blocks available in most FPGAs. For example, the CARRY4 primitive found in Xilinx FPGAs represents the fast carry logic for a slice, typically used as a delay line element in Time-to-Digital Converters [12].

### III. EXPERIMENTS AND RESULTS

#### A. Test Bench

The SDC monitoring functionality was validated by an implementation in a Xilinx Virtex-5 FPGA linked to a sensitive node. The SUT comes from a 64-bit Linear Feedback Shift Register (LFSR) at the rate of system clock, 400 MHz. The test bench is depicted in Fig. 3. A PLLs_CLKMUX module generates the system clock. While no delay change is injected to the SUT, both $FF_1$ and $FF_2$ sample the same value at the rising edge of the clock signal. Hence, the whole system will operate at its nominal frequency with $\tau_{slack}$ = 128 psec to tolerate the maximum SDC obtained experimentally at TRIUMF [4]. However, other values could be set for different measured SDCs. Notice that the timing slack varies with Process-Voltage-Temperature (PVT), as described in Section C. The ADL was implemented with carry chain blocks and flip-flops available in FPGA slices. The ADL outputs can be provided by the outputs of CARRY4 primitive, which are two types: OR-gates output; and the carry multiplexers (MUXCYs) output. The ADL delay is adjusted by selection of any of these outputs in carry chain. The minimum extracted resolution for each carry logic output in Virtex-5 (65 nm technology) is around 20 psec.

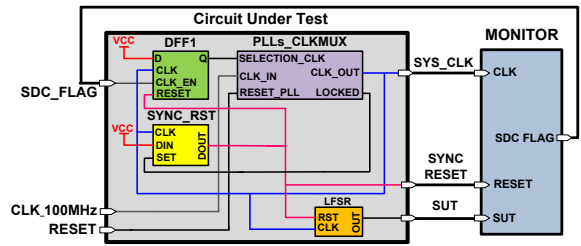

Fig. 3. The test bench configured to validate the delay monitors' functionality.

### B. Experimental Setup

An experimental setup was employed to confirm the functionality of the monitor circuit in real implementation. This setup emulates SEU injection into the FPGA and locates the exact address and position of the fault within the configuration memory frames of FPGA. The experiment benefits from SEU Controller macro (SEUC) embedded in Virtex-5 FPGA [13]. Bit flips are injected through emulation using SEUC via an UART serial communication port. For each bit flip injection, the status of the monitor circuit is captured by the Integrated Logic Analyzer (ILA), ChipScope Pro from Xilinx. The SEUC is set on Detection Only Mode (DOM), which does not perform error correction. For each injected SEU, the SEUC sends back through the UART the exact address location and position of the bit flip (upset) within the configuration memory frames of the FPGA. When the SDC flag is asserted, this address automatically sampled by Chipscope.

Fig. 4 shows the post place and route simulation results of the circuit in Fig. 3. The `sut` signal is fed to $FF_1$ and its delayed version (`delayed_sut`) is fed to $FF_2$ (Fig. 2). At the time of SDC injection on the SUT node (at 576 ns), the signals `dff1` and `dff2` are shifted by the same delay change, hence, the two flip-flops sample different values and the `sdc_flag` will be asserted.

Fig. 5 shows the experimental result obtained by the ILA tool. The UART interface in SEUC will make a brief report that provides the opportunity to record the exact location of any injected bit flip. When SEUC operates in DOM, the faulty address is reported by tracing `uart_tx` signal of SEUC. This signal is the serialized representation of `data_in` parallel input in SEUC transmitter. For each byte transmission, the `Tx_complete` signal is raised while the bytes are sent serially.

Table I presents the overhead and logic utilization by monitor architecture compared to TIMBER flip-flop [7] and scan flip-flop [9] in some comparable features. Similar to any logic circuit, the input capacitive load of monitors will add a delay to the sensitive nodes. This result was not clearly describe in [7, 9]. Delay overhead induced by our monitor on SUT was extracted by Static Timing Analyzer (STA) tool from Xilinx and depends on its physical distance with the SUT. For example, in our test bench, the SUT (Fig. 3) has one fanout in the same slice and its net delay increases from 295 ps to 326 ps, or 10.5 %.

### C. Experiments to Detect Cumulative Delay Change (CDC)

Twelve monitors with different threshold delays ($DC_{Th}$) on the same SUT were used to detect and measure first and cumulative second and third delay changes. Delays and therefore $DC_{Th}$ depend on Process-Voltage-Temperature. Ranges of $DC_{th}$ were extracted from STA for (slow process-0.95V-85°C) and (fast process-1.05V-0°C) corner cases. Each monitor would detect a DC according to its own configured $DC_{th}$ that falls within a delay range. Our experiments on the SUT shown in Fig. 3 allowed to detect 24 first delay changes, 43 second delay changes and 83 third delay changes. For each first delay change, a group of second and third delay changes were detected that were sorted by a tree map. Fig. 6 presents the tree map for one of 24 first delay changes. The fault location for those delay changes are not shown due to the limited space of this abstract. The first column in Fig. 6 falls within a delay range from the best-case $DC_{Th}$ value set in the first monitor, up to the worst-case $DC_{Th}$ value set in the second monitor. The second column shows two of all

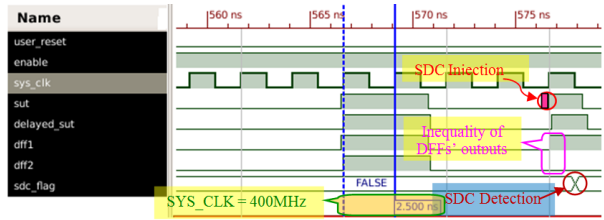

Fig. 4. Post place and route simulation results of SDC detection scheme using SUT-delayed monitor architecture.

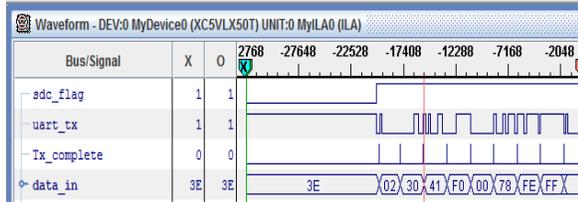

Fig. 5. Experimental result of SDC detection obtained by ChipScope Pro Analyzer (ILA).

TABLE I. Overhead and Device Utilization for Monitor Circuits

| Feature | TIMBER Flip-Flop[7] | Scan Flip-Flop[9] | This Work |
|---|---|---|---|
| Timing Error Coverage | Range of nanoseconds | Range of nanoseconds | $n$ times of ~20 ps |
| Delay Element Type | Inverter chains | OR tree | Carry chain |
| Operational Mode | Low frequency | Mid frequency | High frequency |
| Technology | 90 nm | 90 nm | Virtex-5 (65 nm) |

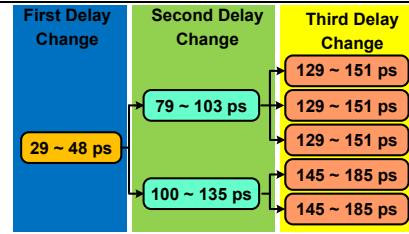

Fig. 6. One set of results for cumulative delay change detection.

43 second delay changes detected by the next monitors. Finally, the third column depicts five of all 83 third delay changes. It is worth mentioning that once a delay change (e.g. 29~48 psec) is detected, this delay change will be added to all monitors linked to the SUT and obviously the cumulative delay change values will be larger. More details about cumulative delay change detection and the corresponding flow graphs for each experiment will be presented extensively at the time of the conference.

## IV. CONCLUSION

Permanent extra combinational delays due to SEU induced by proton radiation in SRAM-based FPGA were observed. The root cause of these delay changes is bit flips of SRAM-cells configuring interconnections and switch boxes, which create shorts between routed and unused wires. As more than 90% of the configuration bits control the routing resources, this paper presented a monitor circuit for delay change detection to be used in any mitigation technique. Post place and route validation results and experimental results confirm the functionality of the monitor. The proposed monitor compares advantageously to TIMBER and scan flip-flop since it can detect delay changes in the range of $n$ times of 20 psec rather than nanoseconds. The monitor loads the sensitive node with a delay overhead as small as 31 psec in best-case and as large as 40 psec in the worst-case while it is closely connected to the sensitive node. Further experiments employing several monitors linked to the sensitive node were performed to detect cumulative delay changes.

## REFERENCES

[1] A. Lesea, *et al.*, "The rosetta experiment: atmospheric soft error rate testing in differing technology FPGAs," *IEEE Trans. Device Mater. Rel.*, 2005.
[2] P. S. Ostler, *et al.*, "SRAM FPGA reliability analysis for harsh radiation environments," *IEEE Trans. Nucl. Sci.*, 2009.
[3] M. Caffrey, *et al.*, "Single-event upsets in SRAM FPGAs," in *Proc. Int. Conf. on Military and Aerospace Programmable Logic Devices*, 2002.
[4] C. Thibeault, *et al.*, "On extra combinational delays in SRAM FPGAs due to transient ionizing radiations," *IEEE Trans. Nucl. Sci.*, 2012.
[5] F. Z. Tazi, *et al.*, "On Extra Delays Affecting I/O Blocks of an SRAM-Based FPGA Due to Ionizing Radiation," *IEEE Trans. Nucl. Sci.*, 2014.
[6] M. Darvishi, *et al.*, "Circuit Level Modeling of Extra Combinational Delays in SRAM-Based FPGAs Due to Transient Ionizing Radiation," *IEEE Trans. Nucl. Sci.*, 2014.
[7] M. R. Choudhury, *et al.*, "Time-borrowing circuit designs and hardware prototyping for timing error resilience," *IEEE Trans. Comput.*, 2014.
[8] M. Nicolaidis, "Time redundancy based soft-error tolerance to rescue nanometer technologies," in *VLSI Test Symp.,Proc.. 17th IEEE*, 1999.
[9] S. Valadimas, *et al.*, "The Time Dilation Technique for Timing Error Tolerance," *IEEE Trans. Comput.*, 2014.
[10] A. Panteloukas, *et al.*, "Timing-resilient Network-on-Chip architectures," in *IOLTS*, 2015.
[11] Xilinx. Virtex-5 FPGA User Guide [UG190].
[12] C. Favi, *et al.*, "A 17ps time-to-digital converter implemented in 65nm FPGA technology," *ACM/SIGDA int. symp. on FPGAs*, 2009.
[13] J. L. Nunes, *et al.*, "Evaluating xilinx SEU controller macro for fault injection," *DSN, 43rd Annual IEEE/IFIP Int. Conf.*, 2013.